%% file: clipper.tex
\pdfoutput=1
\documentclass[letterpaper,twocolumn,10pt]{article}

\usepackage[top=1in, bottom=1in, left=1in, right=1in]{geometry}
\usepackage{usenix}

\usepackage{mathptmx}
\usepackage[scaled=.90]{helvet}
\usepackage{courier}
\usepackage{graphicx}
\usepackage{color}
\usepackage{balance}
\usepackage[algoruled,vlined]{algorithm2e}
\usepackage{amsmath,amsfonts,amssymb}
\usepackage{listings}
\usepackage{tabularx}
\usepackage{multicol,zi4}
\usepackage{float}
\usepackage{subfig}
\usepackage[font=small, labelfont=bf, justification=justified]{caption}
\usepackage{courier,cite}
\usepackage{alltt}
\usepackage{wrapfig}
\usepackage[mmddyyyy]{datetime}

\usepackage[nostamp]{draftwatermark}

\SetWatermarkText{\shortstack{%
  DRAFT \today\\
  DO NOT DISTRIBUTE}}
\SetWatermarkScale{0.18}
\SetWatermarkAngle{90}
\SetWatermarkHorCenter{0.07\paperwidth}

\usepackage[compact]{title sec}

\usepackage[
  pageanchor=true,
  plainpages=false,
  pdfpagelabels,
  bookmarks,
  bookmarksnumbered,
  pdfborder=0 0 0,  
]{hyperref}

\usepackage[T1]{fontenc}

\usepackage[protrusion=true,expansion=true]{microtype}
\graphicspath{{figs/}}

\definecolor{codegreen}{rgb}{0,0.6,0}
\definecolor{codegray}{rgb}{0.5,0.5,0.5}
\definecolor{codepurple}{rgb}{0.58,0,0.82}
\definecolor{backcolour}{rgb}{0.95,0.95,0.92}

\newcommand{\system}{Clipper\xspace}

\newcommand{\topic}[1]{}
\newcommand{\xin}[1]{}
\newcommand{\fixme}[1]{}
\newcommand{\giulio}[1]{}

\newcommand{\eg}{{e.g.,}~}

\newenvironment{myitemize}
{
   \vspace{0mm}
    \begin{list}{$\bullet$ }{}
        \setlength{\topsep}{0em}
        \setlength{\parskip}{0pt}
        \setlength{\partopsep}{0pt}
        \setlength{\parsep}{0pt}
        \setlength{\itemsep}{1mm}
}
{
    \end{list}
}

\newcommand{\term}[1]{\textbf{#1}}

\newcommand{\tableref}[1]{Table~\ref{#1}}
\newcommand{\figref}[1]{Figure~\ref{#1}}
\newcommand{\listref}[1]{Listing~\ref{#1}}

\newcommand{\secref}[1]{\S\ref{#1}}

\begin{document}
\frenchspacing

\title{Clipper: A Low-Latency Online Prediction Serving System}
\author{
{\rm
  Daniel Crankshaw\textsuperscript{*},
  Xin Wang\textsuperscript{*},
  Giulio Zhou\textsuperscript{*}
}\\
{\rm
Michael J. Franklin\textsuperscript{*\textdagger},
Joseph E. Gonzalez\textsuperscript{*},
Ion Stoica\textsuperscript{*}
}\\
\normalsize{\textsuperscript{*}UC Berkeley \qquad
\textsuperscript{\textdagger}The University of Chicago}
}


\maketitle
\thispagestyle{empty}




\input{abstract}

\input{introduction}


\input{challenges}

\input{system}

\input{model-abstraction-layer}

\input{model-selection-layer}


\input{evaluation}

\input{related-work}
\input{conclusion}

\section*{Acknowledgments}

We would like to thank Peter Bailis, Alexey Tumanov, Noah Fiedel, Chris Olston, our shepherd Mike Dahlin, and the anonymous reviewers for their feedback.
This research is supported in part by DHS Award HSHQDC-16-3-00083, DOE Award SN10040 DE-SC0012463, NSF CISE Expeditions Award CCF-1139158, and gifts from Ant Financial, Amazon Web Services, CapitalOne, Ericsson, GE, Google, Huawei, Intel, IBM, Microsoft and VMware.

\newpage

%
\bibliographystyle{abbrv}

{\small
\bibliography{references}
}
%
%
\end{document}

%% file: abstract.tex

\begin{abstract}
Machine learning is being deployed in a growing number of applications which demand real-time, accurate, and robust predictions under heavy query load.
However, most machine learning frameworks and systems only address model training and not deployment.

In this paper, we introduce \system, a 
general-purpose low-latency prediction serving system. 
Interposing between end-user applications and a wide range of machine learning frameworks, Clipper introduces a modular architecture to simplify model deployment across frameworks and applications.
Furthermore, by introducing caching, batching, and adaptive model selection techniques, Clipper reduces prediction latency and improves prediction throughput, accuracy, and robustness without modifying the underlying machine learning frameworks.
We evaluate Clipper on four common machine learning benchmark datasets and demonstrate its ability to meet the latency, accuracy, and throughput demands of online serving applications.
Finally, we compare Clipper to the Tensorflow Serving system and demonstrate that we are able to achieve comparable throughput and latency while enabling model composition and online learning to improve accuracy and render more robust predictions.
\end{abstract}






%

%% file: introduction.tex

\section{Introduction}

The past few years have seen an explosion of applications driven by machine learning, including
recommendation systems~\cite{h20,dato}, voice assistants~\cite{siri,cortana,googlenow}, and ad-targeting~\cite{Graepel10,Agarwal14}.
These applications depend on two stages of machine learning: \emph{training} and \emph{inference}.
Training is the process of building a model from data (\eg movie ratings).
Inference is the process of using the model to make a prediction given an input (e.g., predict a user's rating for a movie).
While training is often computationally expensive, requiring multiple passes over potentially large datasets, inference is often assumed to be inexpensive.
Conversely, while it is acceptable for training to take hours to days to complete, inference must run in real-time,
often on orders of magnitude more queries than during training, 
and is typically part of user-facing applications.

For example, consider an online news organization that wants to deploy a content recommendation service to personalize the presentation of content.
Ideally, the service should be able to recommend articles at interactive latencies (<100ms)~\cite{Yun:2015}, scale to large and growing user populations, sustain the throughput demands of flash crowds driven by breaking news, and provide accurate predictions as the news cycle and reader interests evolve.

The challenges of developing these services differ between the training and inference stages.
On the training side,
developers must choose from a bewildering array of machine learning frameworks with diverse APIs, models, algorithms, and hardware requirements.
Furthermore, they may often need to migrate between models and frameworks as new, more accurate techniques are developed.
Once trained, models must be \emph{deployed} to a prediction serving system to provide low-latency predictions at scale.

Unlike model development, which is supported by sophisticated infrastructure, theory, and systems, model deployment and prediction-serving have received relatively little attention.
Developers must cobble together the necessary pieces from various systems components, and must integrate and support inference across multiple, evolving frameworks, all while coping with ever-increasing demands for scalability and responsiveness.
As a result, the deployment, optimization, and maintenance of machine learning services is difficult and error-prone.


\begin{figure}[t]
\centering
\includegraphics[width=0.90\linewidth]{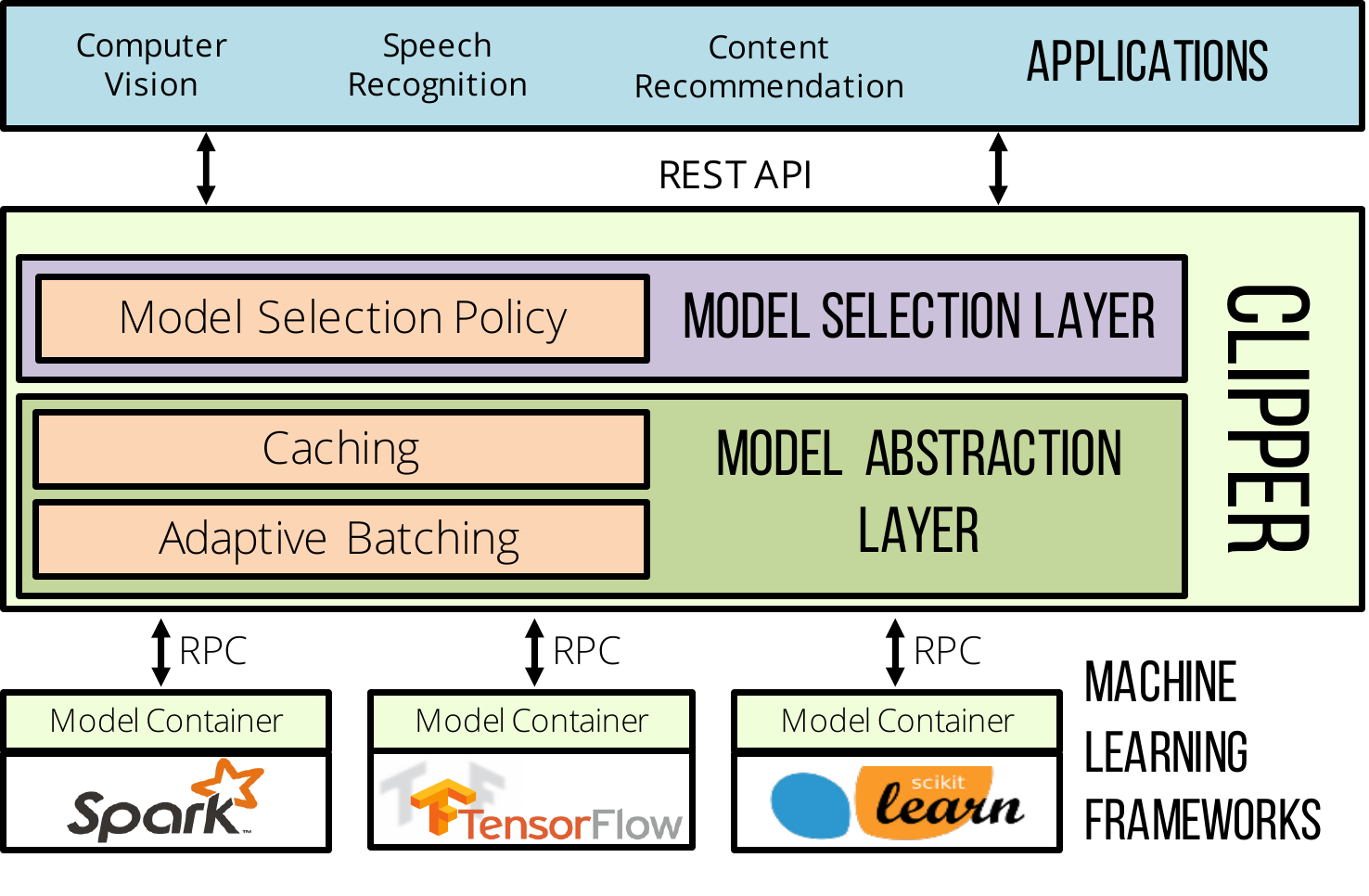}
\vspace{-1em}
\caption{\small \textbf{The \system Architecture.}
\vspace{-2em}
}
\label{fig:sysarch}
\end{figure}

To address these challenges, we propose \system, a \emph{layered architecture} system (\figref{fig:sysarch}) that reduces the complexity of implementing a prediction serving stack and achieves three crucial properties of a prediction serving system: \emph{low latencies}, \emph{high throughputs}, and \emph{improved accuracy}.
Clipper is divided into two layers: (1) the model abstraction layer, and (2) the model selection layer. The first layer exposes a common API that abstracts away the heterogeneity of existing ML frameworks and models. Consequently, models can be modified or swapped transparently to the application.
The model selection layer sits above the model abstraction layer and dynamically selects and combines predictions across competing models to provide more accurate and robust predictions.




To achieve low latency, high throughput predictions, Clipper implements a range of optimizations.
In the model abstraction layer, Clipper caches predictions on a per-model basis and implements adaptive batching to maximize throughput given a query latency target.
In the model selection layer, Clipper implements techniques to improve prediction accuracy and latency.
To improve accuracy, Clipper exploits bandit and ensemble methods to robustly select and combine predictions from multiple models and estimate prediction uncertainty.
In addition, Clipper is able to adapt the model selection independently for each user or session.
To improve latency, the model selection layer adopts a straggler mitigation technique to render predictions without waiting for slow models. 
Because of this layered design, neither the end-user applications nor the underlying machine learning frameworks need to be modified to take advantage of these optimizations.

We implemented \system in Rust and added support for several of the most widely used machine learning frameworks: Apache Spark MLLib~\cite{mllib}, Scikit-Learn~\cite{scikitlearn}, Caffe~\cite{jia2014caffe}, TensorFlow~\cite{abaditensorflow}, and HTK~\cite{htkbook}.
While these frameworks span multiple application domains, programming languages, and system requirements, each was added using fewer than 25 lines of code.

We evaluate \system using four common machine learning benchmark datasets and demonstrate that \system is able to render low and bounded latency predictions (<20ms), scale to many deployed models even across machines, quickly select and adapt the best combination of models, and dynamically trade-off accuracy and latency under heavy query load.
We compare \system to the Google TensorFlow Serving system~\cite{tfserving},
an industrial grade prediction serving system tightly integrated with the TensorFlow training framework.
We demonstrate that \system's modular design and broad functionality impose minimal performance cost, achieving comparable prediction throughput and latency to TensorFlow Serving while supporting substantially more functionality.
In summary, our key contributions are:
\begin{myitemize}
\item A layered architecture that abstracts away the complexity associated with serving predictions in existing machine learning frameworks (\secref{sec:system}).
\item A set of novel techniques to reduce and bound latency while maximizing throughput that generalize across machine learning frameworks (\secref{sec:model-abstraction}).
\item A model selection layer that enables online model selection and composition to provide robust and accurate predictions for interactive applications (\secref{sec:selection-layer}).

\end{myitemize}


%% file: challenges.tex

\section{Applications and Challenges}
\label{sec:challenges}

\begin{figure}[t]
\centering
\includegraphics[width=0.95\linewidth]{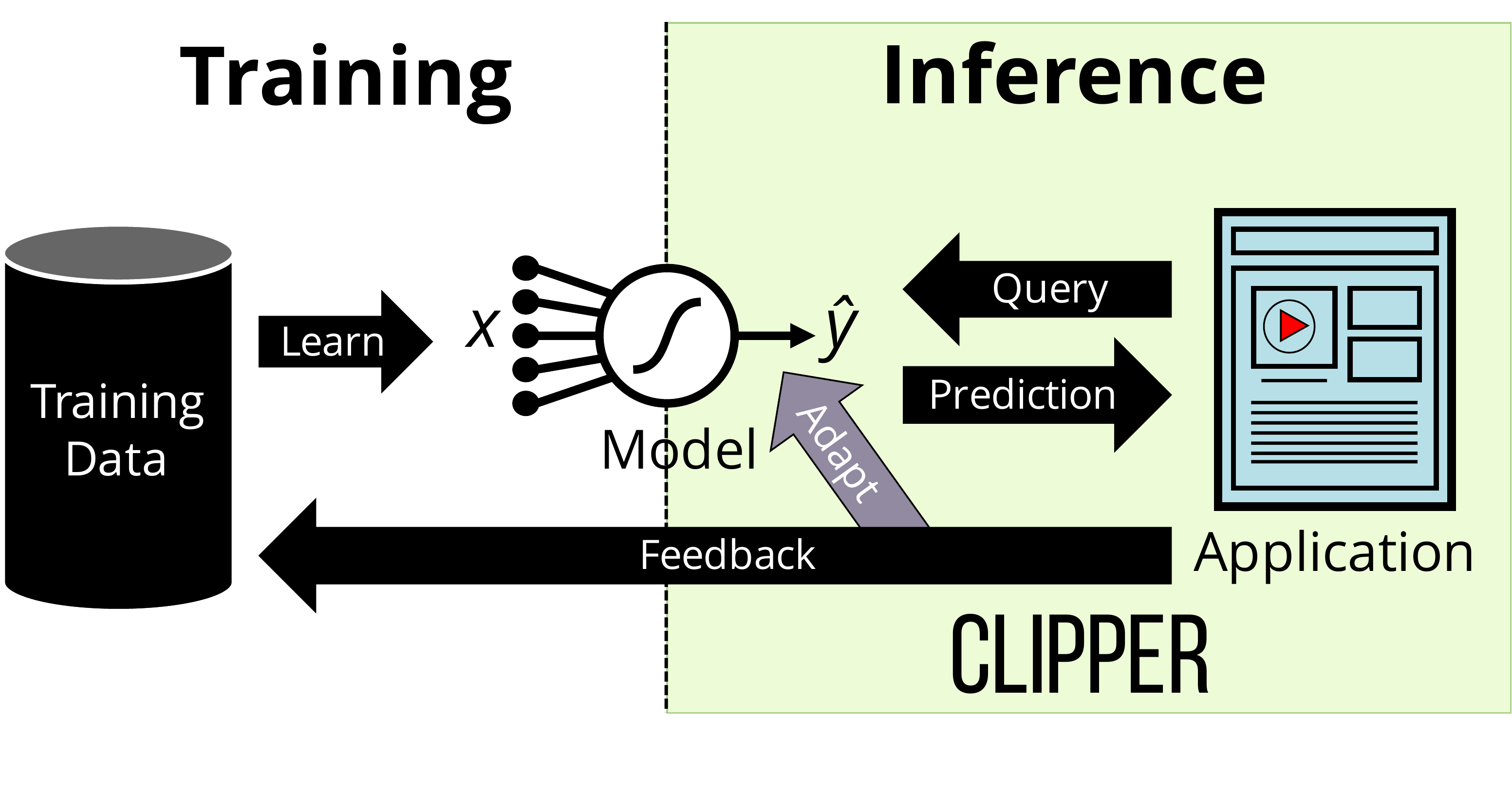}
\vspace{-1em}
\caption{\small \textbf{Machine Learning Lifecycle.}}
\label{fig:big_picture}
\vspace{-0.5em}
\end{figure}

The machine learning life-cycle (\figref{fig:big_picture}) can be divided into two distinct phases: \emph{training} and \emph{inference}.
Training is the process of estimating a model from data.
Training is often computationally expensive requiring multiple passes over large datasets and can take hours or even days to complete~\cite{Chen16,visualattention,resnet}.
Much of the innovation in systems for machine learning has focused on model training with the development of systems like Apache Spark~\cite{spark}, the Parameter Server~\cite{MuLi}, PowerGraph~\cite{graphlab}, and Adam~\cite{Adam}.


A wide range of machine learning frameworks have been developed to address the challenges of training.
Many specialize in particular models such as TensorFlow~\cite{abaditensorflow} for deep learning or Vowpal Wabbit~\cite{vowpal} for large linear models. Others are specialized for specific application domains such as Caffe~\cite{jia2014caffe} for computer vision or HTK~\cite{htkbook} for speech recognition.
Typically, these frameworks leverage advances in parallel and distributed systems to scale the training process.

Inference is the process of evaluating a model to render predictions. 
In contrast to training, inference does not involve complex iterative algorithms and is therefore generally assumed to be easy.
As a consequence, there is little research studying the process of inference and most machine learning frameworks provide only basic support for offline batch inference -- often with the singular goal of evaluating the model training algorithm.
However, scalable, accurate, and reliable inference presents fundamental system challenges that will likely dominate the challenges of training as machine learning adoption increases.
In this paper we focus on the less studied but increasingly important challenges of \emph{inference}.


\subsection{Application Workloads}
\label{subsec:appworkloads}

To illustrate the challenges of inference and provide a benchmark on which to evaluate \system, we describe two canonical real-world applications of machine learning: \emph{object recognition} and \emph{speech recognition}.

\subsubsection*{Object Recognition}
Advances in deep learning have spurred rapid progress in computer vision, especially in object recognition problems -- the task of identifying and labeling the objects in a picture.
Object recognition models form an important building block in many computer vision applications ranging from image search to self-driving cars.

As users interact with these applications, they provide feedback about the accuracy of the predictions, either by explicitly labeling images (\eg tagging a user in an image) or implicitly by indicating whether the provided prediction was correct (\eg clicking on a suggested image in a search).
Incorporating this feedback quickly can be essential to eliminating failing models and providing a more personalized experience for users.

\textbf{Benchmark Applications:}
We use the well studied MNIST~\cite{mnist}, CIFAR-10~\cite{cifardata}, and ImageNet~\cite{imagenet} datasets to evaluate increasingly difficult object recognition tasks with correspondingly larger inputs.
For each dataset, the prediction task requires identifying the correct label for an image based on its pixel values. MNIST is a common baseline dataset used to demonstrate the potential of a new algorithm or technique, and both deep learning and more classical machine learning models perform well on MNIST. On the other hand, for CIFAR-10 and Imagenet, deep learning significantly outperforms other methods.
By using three different datasets, we evaluate \system's performance when serving models that have a wide variety of computational requirements and accuracies.

\subsubsection*{Automatic Speech Recognition}
Another successful application of machine learning is automatic speech recognition.
A speech recognition model is a function from a spoken audio signal to the corresponding sequence of words.
Speech recognition models can be relatively large~\cite{chelba2012large} and are often composed of many complex sub-models trained using specialized speech recognition frameworks (\eg HTK~\cite{htkbook}).
Speech recognition models are also often personalized to individual users to accommodate variations in dialect and accent.

In most applications, inference is done online as the user speaks.
Providing real-time predictions is essential to user experience~\cite{agarwal2009matchmaking} and enables new applications like real-time translation~\cite{skype}.
However, inference in speech models can be costly~\cite{chelba2012large} requiring the evaluation of large tensor products in convolutional neural networks.

As users interact with speech services, they provide implicit signal about the quality of the speech predictions which can be used to identify the dialect.
Incorporating this feedback quickly improves user experience by allowing us to choose models specialized for a user's dialect.

\textbf{Benchmark Application:}
To evaluate the benefit of personalization and online model-selection on a dataset with real user data, we built a speech recognition service with the widely used TIMIT speech corpus~\cite{timit} and the HTK~\cite{htkbook} machine learning framework.
This dataset consists of voice recordings for 630 speakers in eight dialects of English.
We randomly drew users from the test corpus and simulated their interaction with our speech recognition service using their pre-recorded speech data.

\subsection{Challenges}

Motivated by the above applications, we outline the key challenges of prediction serving and describe how \system addresses these challenges.

\subsubsection*{Complexity of Deploying Machine Learning}

There is a large and growing number of machine learning frameworks~\cite{collobert2011torch7,jia2014caffe,abaditensorflow,chen2015mxnet,theano}.
Each framework has strengths and weaknesses and many are optimized for specific models or application domains (\eg computer vision).
Thus, there is no dominant framework and often multiple frameworks may be used for a single application (\eg speech recognition and computer vision in automatic captioning).
Furthermore, machine learning is an iterative process and the best framework may change as an application evolves over time (\eg as a training dataset grows to require distributed model training).
Although common model exchange formats have been proposed~\cite{pmml,pfa}, they have never achieved widespread adoption
because of the rapid and fundamental changes in state-of-the-art techniques and additional source of errors from parallel implementations for training and serving.
Finally, machine learning frameworks are often developed by and for machine learning experts and are therefore heavily optimized towards model development rather than deployment.
As a consequence of these design decisions, application developers are forced to accept reduced accuracy by forgoing the use of a model well-suited to the task or to incur the substantially increased complexity of integrating and supporting multiple machine learning frameworks.

\textbf{Solution:}  \system introduces a model abstraction layer and common prediction interface that isolates applications from variability in machine learning frameworks (\secref{sec:model-abstraction}) and simplifies the process of deploying a new model or framework to a running application.

\subsubsection*{Prediction Latency and Throughput}

The \emph{prediction latency} is the time it takes to render a prediction given a query.
Because prediction serving is often on the critical path, predictions must both be fast and have bounded tail latencies to meet service level objectives~\cite{Yun:2015}.
While simple linear models are fast, more sophisticated and often more accurate models such as support vector machines, random forests, and deep neural networks are much more computationally intensive and can have substantial latencies (50-100ms)~\cite{chen2015mxnet} (see \figref{fig:tf_comp} for details).
In many cases accuracy can be improved by combining models but at the expense of stragglers and increased tail latencies.
Finally, most machine learning frameworks are optimized for offline batch processing and not single-input prediction latency.
Moreover, the low and bounded latency demands of interactive applications are often at odds with
the design goals of machine learning frameworks.




The computational cost of sophisticated models can substantially impact prediction throughput.
For example, a relatively fast neural network which is able to render 100 predictions per second is still orders of magnitude slower than a modern web-server.
While batching prediction requests can substantially improve throughput by exploiting optimized BLAS libraries, SIMD instructions, and GPU acceleration it can also adversely affect prediction latency.
Finally, under heavy query load it is often preferable to marginally degrade accuracy rather than substantially increase latency or lose availability~\cite{Ganjam:2015tq,Agarwal14}.

\textbf{Solution:} \system automatically and adaptively batches prediction requests to maximize the use of batch-oriented system optimizations in machine learning frameworks while ensuring that prediction latency
objectives are still met (\secref{subsec:batching}).
In addition, \system employs straggler mitigation techniques to reduce and bound tail latency, enabling model developers to experiment with complex models without affecting serving latency (\secref{subsec:straggler-mitigation}).

\subsubsection*{Model Selection}

Model development is an iterative process producing many models reflecting different feature representations, modeling assumptions, and machine learning frameworks.
Typically developers must decide which of these models to deploy based on offline evaluation using stale datasets or engage in costly online A/B testing.
When predictions can influence future queries (\eg content recommendation), offline evaluation techniques can be heavily biased by previous modeling results.
Alternatively, A/B testing techniques\cite{Agarwal} have been shown~
to be statistically inefficient --- requiring data to grow \emph{exponentially} in the number of candidate models.
The resulting choice of model is typically static and therefore susceptible to changes in model performance due to factors such as feature corruption or concept drift\cite{Sculley2015}.
In some cases the best model may differ depending on the context (\eg user or region) in which the query originated.
Finally, predictions from more than one model can often be combined in ensembles to boost prediction accuracy and provide more robust predictions with confidence bounds.

\textbf{Solution:} \system leverages adaptive online model selection and ensembling techniques to incorporate feedback and automatically select and combine predictions from models that can span multiple machine learning frameworks.

\subsection{Experimental Setup}

Because we include microbenchmarks of many of \system's features as we introduce them, we present the experimental setup now.
For each of the three object recognition benchmarks, the prediction task is predicting the correct label given the raw pixels of an unlabeled image as input. We used a variety of models on each of the object recognition benchmarks.
For the speech recognition benchmark, the prediction task is predicting the phonetic transcription of the raw audio signal.
For this benchmark, we used the HTK Speech Recognition Toolkit~\cite{htkbook} to learn Hidden Markov Models whose outputs are sequences of phonemes representing the transcription of the sound.
Details about each dataset are presented in~\tableref{tab:datasets}.


\begin{table}[t]
\centering
\small

    \begin{tabular}[b]{ | l | l | l | c | c | c |  }
      \hline
      \textbf{Dataset} & \textbf{Type} & \textbf{Size} & \textbf{Features} & \textbf{Labels} \\
      \hline
      MNIST \cite{mnist} & Image & 70K & 28x28 & 10 \\
      CIFAR \cite{cifardata} & Image & 60k & 32x32x3 & 10 \\
      ImageNet \cite{imagenet} & Image & ~1.26M & 299x299x3 & 1000  \\
      Speech \cite{timit} & Sound & 6300 & 5 sec. & 39  \\
      \hline
   \end{tabular}
  \vspace{-4mm}
  \caption{\small \textbf{Datasets.} The collection of real-world benchmark datasets used in the experiments.}
    \vspace{-5mm}
    \label{tab:datasets}
\end{table}

Unless otherwise noted, all experiments were conducted on a single server.
All machines used in the experiments
contain 2 Intel Haswell-EP CPUs and 256 GB of RAM running Ubuntu 14.04 on Linux 4.2.0.
TensorFlow models were executed on a Nvidia Tesla K20c GPUs with 5 GB of GPU memory and 2496 cores.
In the scaling experiment presented in~\figref{fig:gpu-scaling}, the servers in the cluster were connected with both a 10Gbps and 1Gbps network. For each network, all the servers were located on the same switch.
Both network configurations were investigated.

%% file: system.tex

\section{System Architecture}
\label{sec:system}

%


\system is divided into \emph{model selection} and \emph{model abstraction} layers (see \figref{fig:sysarch}).
The model abstraction layer is responsible for providing a common prediction interface, ensuring resource isolation, and optimizing the query workload for batch oriented machine learning frameworks.
The model selection layer is responsible for dispatching queries to one or more models and combining their predictions based on feedback to improve accuracy, estimate uncertainty, and provide robust predictions.




Before presenting the detailed \system system design we first describe the path of a prediction request through the system.
Applications issue prediction requests to \system through application facing REST or RPC APIs.
Prediction requests are first processed by the model selection layer.
Based on properties of the prediction request and recent feedback, the model selection layer dispatches the prediction request to one or more of the  models through the model abstraction layer.


The model abstraction layer first checks the prediction cache for the query before assigning the query to an adaptive batching queue associated with the desired model.
The adaptive batching queue constructs batches of queries that are tuned for the machine learning framework and model.
A cross language RPC is used to send the batch of queries to a model container hosting the model in its native machine learning framework.
To simplify deployment, we host each model container in a separate Docker container.
After evaluating the model on the batch of queries, the predictions are sent back to the model abstraction layer which populates the prediction cache and returns the results to the model selection layer.
The model selection layer then combines one or more of the predictions to render a final prediction and confidence estimate.  
The prediction and confidence estimate are then returned to the end-user application.

Any feedback the application collects about the quality of the predictions is sent back to the model selection layer through the same application-facing REST/RPC interface.
The model selection layer joins this feedback with the corresponding predictions to improve how it selects and combines future predictions.

We now present the model abstraction layer and the model selection layer in greater detail.

%% file: model-abstraction-layer.tex

\section{Model Abstraction Layer }
\label{sec:model-abstraction}

The \term{Model Abstraction Layer} (\figref{fig:sysarch}) provides a common interface across machine learning frameworks.
It is composed of a prediction cache, an adaptive query-batching component, and a set of model containers connected to \system via a lightweight RPC system.
This modular architecture enables caching and batching mechanisms to be shared across frameworks while also scaling to many concurrent models and simplifying the addition of new frameworks.

\subsection{Overview}

At the top of the model abstraction layer is the prediction cache (\secref{subsec:caching}).
The prediction caches provides a partial pre-materialization mechanism for frequent queries and accelerates the adaptive model selection techniques described in \secref{sec:selection-layer} by enabling efficient joins between recent predictions and feedback.




The batching component (\secref{subsec:batching}) sits below the prediction cache and aggregates point queries into mini-batches that are dynamically resized for each model container to maximize throughput.
Once a mini-batch is constructed for a given model it is dispatched via the RPC system to the container for evaluation.

Models deployed in \system are each encapsulated within their own lightweight container (\secref{sec:model-containers}), communicating with \system through an RPC mechanism that provides a uniform interface to \system and simplifies the deployment of new models.
The lightweight RPC system minimizes the overhead of the container-based architecture and simplifies cross-language integration.



In the following sections we describe each of these components in greater detail and discuss some of the key algorithmic innovations associated with each.

\subsection{Caching}
\label{subsec:caching}
For many applications (\eg content recommendation), predictions concerning popular items are requested frequently.
By maintaining a prediction cache, \system can serve these frequent queries without evaluating the model.
This substantially reduces latency and system load by eliminating the additional cost of model evaluation. 

In addition, caching in Clipper serves an important role in model selection (\secref{sec:selection-layer}). To select models intelligently \system needs to join the original predictions with any feedback it receives.
Since feedback is likely to return soon after predictions are rendered~\cite{McMahan:2013cq},
even infrequent or unique queries can benefit from caching.

For example, even with a small ensemble of four models (a random forest, logistic regression model, and linear SVM trained in Scikit-Learn and a linear SVM trained in Spark), prediction caching increased feedback processing throughput in \system by 1.6x from roughly 6K to 11K observations per second.
%


The prediction cache acts as a function cache for the generic prediction function:
\vspace{-.3em}
\begin{equation*}
    \texttt{Predict(m: ModelId, x: X) -> y: Y}
\vspace{-.3em}
\end{equation*}
that takes a model id $m$ along with the query $x$ and computes the corresponding model prediction $y$.
The cache exposes a simple non-blocking \emph{request} and \emph{fetch} API.
When a prediction is needed, the \emph{request} function is invoked which notifies the cache to compute the prediction if it is not already present and returns a boolean indicating whether the entry is in the cache.
The \emph{fetch} function checks the cache and returns the query result if present.


\system employs an LRU eviction policy for the prediction cache, using the standard CLOCK~\cite{clockcacheeviction} cache eviction algorithm.
With an adequately sized cache, frequent queries will not be evicted and the cache serves as a partial pre-materialization mechanism for hot items.
However, because adaptive model selection occurs \emph{above the cache} in \system, changes in predictions due to model selection do not invalidate cache entries.


\subsection{Batching}
\label{subsec:batching}

\begin{figure*}[ht]
\centering \small
\includegraphics[width=\linewidth]{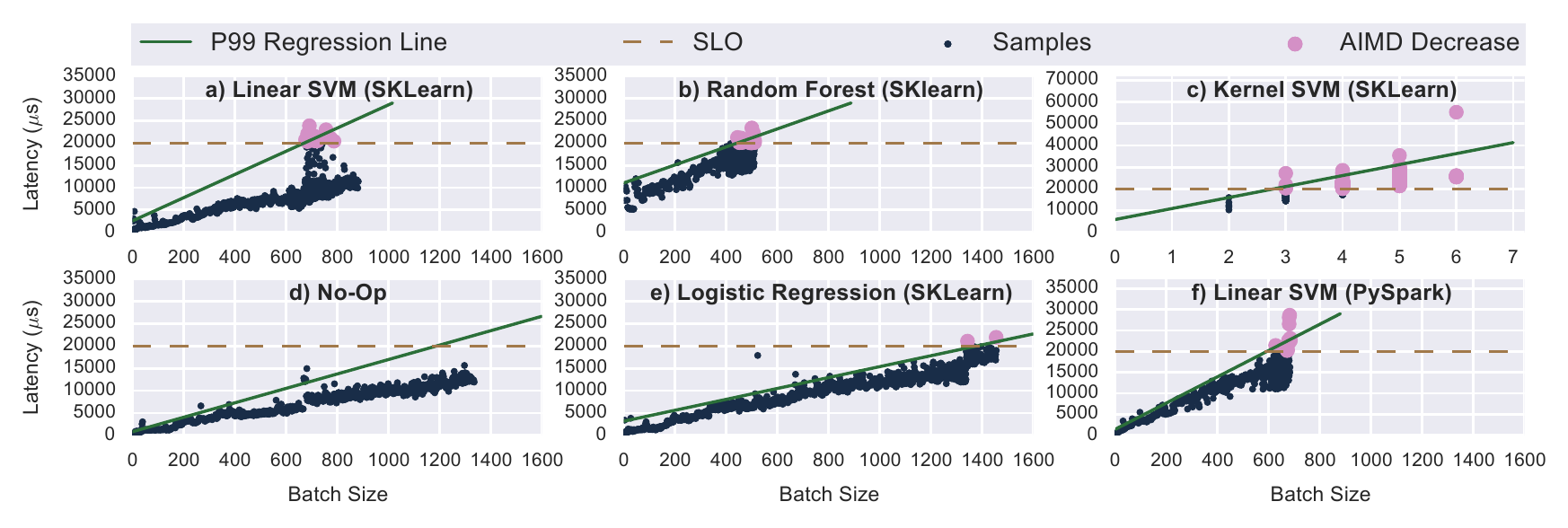}
\vspace{-2em}
\caption{\small \textbf{Model Container Latency Profiles.} We measured the batching latency profile of several models trained on the MNIST
  benchmark dataset. The models were trained using Scikit-Learn (SKLearn) or Spark and were chosen to represent several of the most widely used types of models. The No-Op Container measures the system overhead of the model containers and RPC system.
}
\label{fig:batching-scatterplots}
\vspace{-1em}
\end{figure*}

\begin{figure}[t]
\centering \small
\includegraphics[width=0.95\columnwidth]{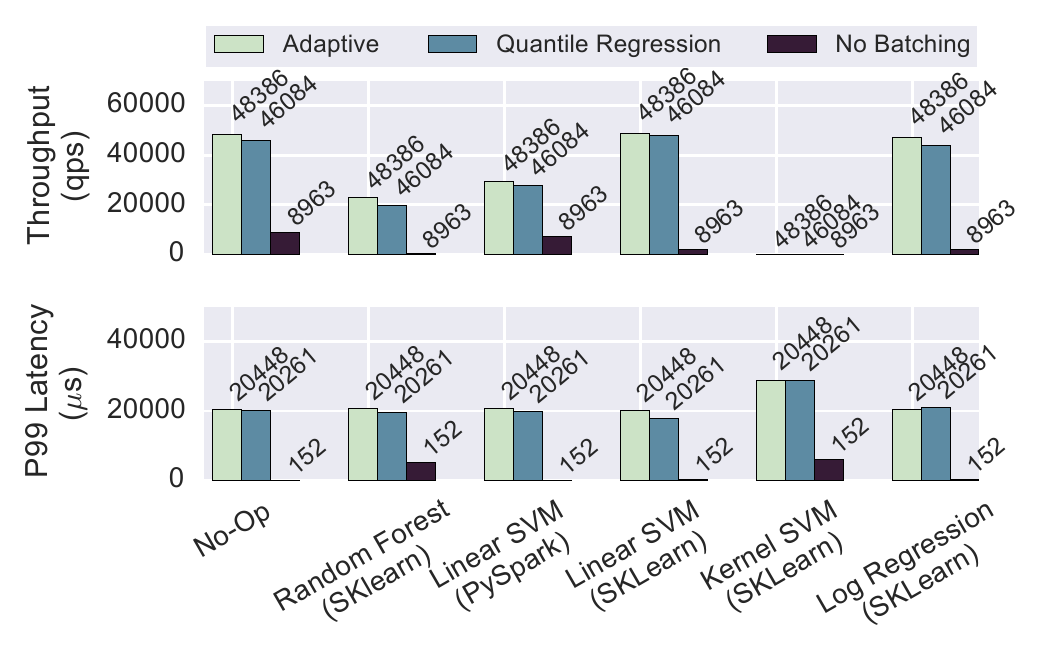}
\vspace{-1.3em}
\caption{\small \textbf{Comparison of Dynamic Batching Strategies.}
  %
}
\label{fig:batch-strat-comp}
\vspace{-1.3em}
\end{figure}

The \system batching component transforms the concurrent stream of prediction queries received by \system into batches that more closely match the workload assumptions made by machine learning frameworks while simultaneusly amortizing RPC and system overheads.
Batching improves throughput and utilization of often costly physical resources such as GPUs, but it does so at the expense of increased latency by requiring all queries in the batch to complete before returning a single prediction.

We exploit an explicitly stated latency service level objective (SLO) to \emph{increase latency} in exchange for substantially improved throughput.
By allowing users to specify a latency objective, \system is able to tune batched query evaluation to maximize throughput while still meeting the latency requirements of interactive applications.
For example, requesting predictions in sufficiently large batches can improve throughput by up to 26x (the Scikit-Learn SVM in \figref{fig:batch-strat-comp})
while meeting a 20ms latency SLO.

Batching increases throughput via two mechanisms.
First, batching amortizes the cost of RPC calls and internal framework overheads such as copying inputs to GPU memory.
Second, batching enables machine learning frameworks to exploit existing data-parallel optimizations by performing batch inference on many inputs simultaneously (\eg by using the GPU or BLAS acceleration).

As the model selection layer dispatches queries for model evaluation, they are placed on queues associated with  model containers.
Each model container has its own adaptive batching queue tuned to the latency profile of that container and a corresponding thread to process predictions.
Predictions are processed in batches by
removing as many queries as possible from a queue up to the maximum batch size \emph{for that model container} and sending the queries as a single batch prediction RPC to the container for evaluation.
\system imposes a \emph{maximum} batch size to ensure that latency objectives are met and avoid excessively delaying the first queries in the batch.



Frameworks that leverage GPU acceleration such as TensorFlow often enforce static batch sizes to maintain a consistent data layout across evaluations of the model.
These frameworks typically encode the batch size directly into the model definition in order to fully exploit GPU parallelism.
When rendering fewer predictions than the batch size, the input must be padded to reach the defined size, reducing model throughput without any improvement in prediction latency.
Careful tuning of the batch size should be done to maximize inference performance, but this tuning must be done offline and is fixed by the time a model is deployed.

However, most machine learning frameworks can efficiently process variable-sized batches at serving time.
Yet differences between the framework implementation and choice of model and inference algorithm can lead to orders of magnitude variation in model throughput and latency.
As a result, the latency profile -- the expected time to evaluate a batch of a given size -- varies substantially between model containers.
For example, in~\figref{fig:batching-scatterplots} we see that the maximum batch size that can be executed within a 20ms latency SLO differs by 241x between the linear SVM which does a very simple vector-vector multiply to perform inference and the kernel SVM which must perform a sequence of expensive nearest-neighbor calculations to evaluate the kernel. As a consequence, the linear SVM can achieve throughput of nearly 30,000 qps while the kernel SVM is limited to 200 qps under this SLO.
Instead of requiring application developers to manually tune the batch size for each new model, \system employs a simple adaptive batching scheme to dynamically find and adapt the maximum batch size.

\subsubsection{Dynamic Batch Size}
\label{sec:adaptbatch}

We define the optimal batch size as the batch size that maximizes throughput subject to the constraint that the batch evaluation latency is under the target SLO.
To automatically find the optimal maximum batch size for each model container we employ an additive-increase-multiplicative-decrease (AIMD) scheme.
Under this scheme, we additively increase the batch size by a fixed amount until the latency to process a batch exceeds the latency objective.
At this point, we perform a small multiplicative backoff, reducing the batch size by 10\%.
Because the optimal batch size does not fluctuate substantially, we use a much smaller backoff constant than other Additive-Increase, Multiplicative-Decrease schemes~\cite{Chiu:1989}.

Early performance measurements (\figref{fig:batching-scatterplots}) suggested a stable linear relationship between batch size and latency across several of the modeling frameworks.
As a result, we also explored the use of quantile regression to estimate the 99th-percentile (P99) latency as a function of batch size and set the maximum batch size accordingly. 
We compared the two approaches on a range of commonly used Spark and Scikit-Learn models in~\figref{fig:batch-strat-comp}.
Both strategies provide significant performance improvements over the baseline strategy of no batching, achieving up to a 26x throughput increase in the case of the Scikit-Learn linear SVM, demonstrating the performance gains that batching provides.
While the two batching strategies perform nearly identically, 
the AIMD scheme is significantly simpler and easier to tune.
Furthermore, the ongoing adaptivity of the AIMD strategy makes it robust to changes in throughput capacity of a model (\eg during a garbage collection pause in Spark).
As a result, \system employs the AIMD scheme as the default.

\subsubsection{Delayed Batching}
Under moderate or bursty loads, the batching queue may contain less queries than the maximum batch size
when the next batch is ready to be dispatched. For some models, briefly delaying the dispatch to allow more queries to arrive can significantly improve throughput under bursty loads. Similar to the motivation for Nagle's algorithm~\cite{Nagle84}, the gain in efficiency is a result of the ratio of the fixed cost for sending a batch to the variable cost of increasing the size of a batch.

\begin{figure}[t]

\centering \small
\includegraphics[width=0.95\columnwidth]{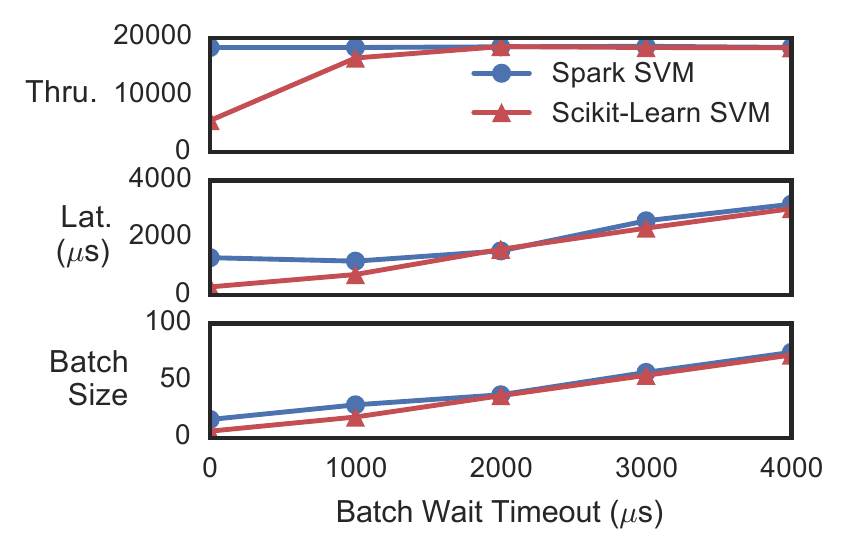}
\vspace{-1.5em}
\caption{\small \textbf{Throughput Increase from Delayed Batching.}}
\label{fig:batch-delay}
\end{figure}

In~\figref{fig:batch-delay}, we compare the gain in efficiency (measured as increased throughput) from delayed batching for two models.
Delayed batching provides no increase in throughput for the Spark SVM because Spark is already relatively efficient at processing small batch sizes and can keep up with the moderate serving workload using batches much smaller than the optimal batch size. In contrast, the Scikit-Learn SVM has a high fixed cost for processing a batch but employs BLAS libraries to do efficient parallel inference on many inputs at once. As a consequence, a 2ms batch delay provides a 3.3x improvement in throughput and allows the Scikit-Learn model container to keep up with
the throughput demand while remaining well below the 10-20ms latency objectives needed for interactive applications.

\subsection{Model Containers}
\label{sec:model-containers}


\begin{figure}[t]
  \lstset{basicstyle=\small\ttfamily, keywords={interface}, frame=tb,
  label=lst:containerapi, captionpos=b, caption={\small \textbf{Common Batch Prediction Interface for Model Containers.}
The batch prediction function is called via the RPC interface to compute the predictions for a batch of inputs. The return type is a nested list because each input may produce multiple outputs.}}
    \begin{lstlisting}
interface Predictor<X,Y> {
  List<List<Y>> pred_batch(List<X> inputs);
}
    \end{lstlisting}
\vspace{-2em}
\end{figure}

Model containers encapsulate the diversity of machine learning frameworks and model implementations within a
uniform ``narrow waist'' remote prediction API.
To add a new type of model to \system, model builders only need to implement the standard batch prediction interface in \listref{lst:containerapi}.
\system includes language specific container bindings for C++, Java, and Python.
The model container implementations for most of the models in this paper only required a few lines of code. 

To achieve process isolation, each model is managed
in a separate Docker container.
By placing models in separate containers, we ensure that variability in performance and stability of relatively immature state-of-the-art machine learning frameworks does not interfere with the overall availability of \system.
Any state associated with a model, such as the model parameters, is provided to the container during initialization and the container itself is stateless after initialization.
As a result, resource intensive machine learning frameworks can be replicated across multiple machines or given access to specialized hardware (\eg GPUs) when needed to meet serving demand.

\subsubsection{Container Replica Scaling}

\system supports replicating model containers, both locally and across a cluster, to improve prediction throughput and leverage additional hardware accelerators.
Because different replicas
can have different performance characteristics, particularly when spread across a cluster, \system performs adaptive batching independently for each replica.

\begin{figure}[t]
\centering \small
\includegraphics[width=0.98\columnwidth]{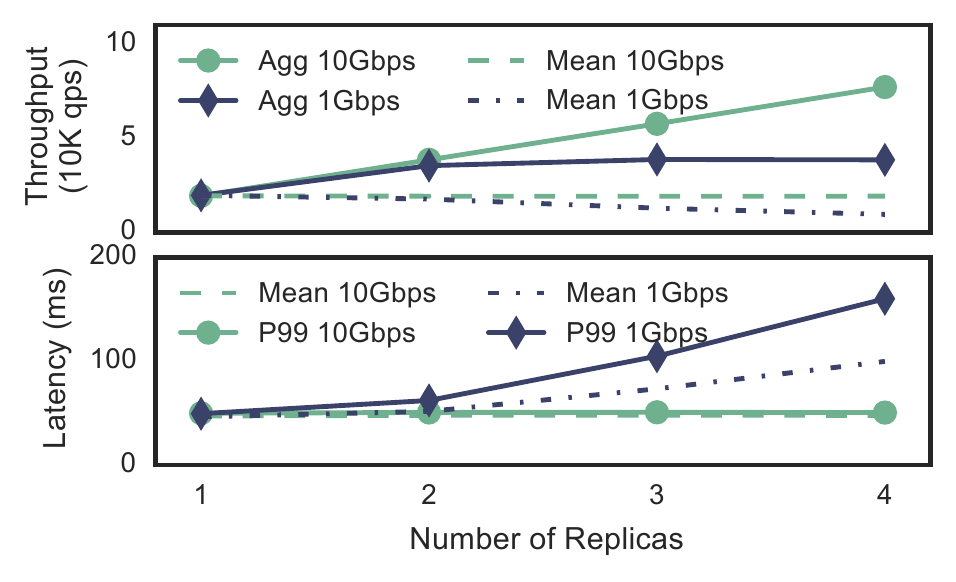}
\vspace{-1em}
\caption{\small \textbf{Scaling the Model Abstraction Layer Across a GPU Cluster.}
The solid lines refer to aggregate throughput of all the model replicas and the dashed lines refer to the mean per-replica throughput.
}
\label{fig:gpu-scaling}
\vspace{-1.7em}
\end{figure}

In~\figref{fig:gpu-scaling} we demonstrate the linear throughput scaling that \system can achieve by replicating model containers across a cluster. With a four-node GPU cluster connected through a 10Gbps Ethernet switch, \system gets a 3.95x throughput increase from 19,500 qps when using a single model container running on a local GPU to 77,000 qps when using four replicas each running on a different machine. Because the model containers in this experiment are computationally intensive and run on the GPU, GPU throughput is the bottleneck and \system's RPC system can easily saturate the GPUs.
However, when the cluster is connected through a 1Gbps switch, the aggregate throughput of the GPUs is higher than 1Gbps and so the network becomes saturated when replicating to a second remote machine.
As machine-learning applications begin to consume increasingly bigger inputs, scaling from handcrafted features to large images, audio signals, or even video, the network will continue to be a bottleneck to scaling out prediction serving applications.
This suggests the need for research into efficient networking strategies for remote predictions on large inputs.

%

%% file: model-selection-layer.tex
\section{Model Selection Layer}
\label{sec:selection-layer}

The \term{Model Selection Layer} uses feedback to dynamically select one or more of the deployed models and combine their outputs to provide more accurate and robust predictions. By allowing many candidate models to be deployed simultaneously and relying on feedback to adaptively determine the best model or combination of models, the model selection layer simplifies the deployment process for new models. By continuously learning from feedback throughout the lifetime of an application, the model selection layer automatically compensates for failing models without human intervention. By combining predictions from multiple models, the model selection layer boosts application accuracy and estimates prediction confidence.

There are a wide range of techniques for model selection and composition that span a tradeoff space of computational overhead and application accuracy.
However, most of these techniques can be expressed with a simple \emph{select}, \emph{combine}, and \emph{observe} API.
We capture this API in the model selection policy interface (\listref{lst:predhandler}) which governs the behavior of the model selection layer and allows users to introduce new model selection techniques themselves.

The model selection policy (\listref{lst:predhandler}) defines four essential functions as well as a few basic types.
In addition to the query and prediction types \emph{X} and \emph{Y}, the state type \emph{S} encodes the learned state of the selection algorithm.
The \emph{init} function returns an initial instance of the selection policy state.
We isolate the selection policy state and require an initialization function to enable \system to efficiently instantiate many instances of the selection policy for fine-grained contextualized model selection (\secref{subsec:contextualization}).
The \emph{select} and \emph{combine} functions
are responsible for choosing which models to query and how to combine the results.
In addition, the \emph{combine} function can compute other information about the predictions. For example, in \secref{subsec:robust-predictions} we leverage the \emph{combine} function to provide a prediction confidence score.
Finally, the \emph{observe} function is used to update the state \emph{S} based on feedback from front-end applications.

\begin{figure}[t]
  \lstset{basicstyle=\small\ttfamily, keywords={interface}, frame=tb,
    label=lst:predhandler, captionpos=b, alsoletter={&},
    caption={\small \textbf{Model Selection Policy Interface.}
    }
}
\begin{lstlisting}
interface SelectionPolicy<S, X, Y> {
  S init();
  List<ModelId> select(S s, X x);
  pair<Y, double> combine(S s, X x,
    Map<ModelId, Y> pred);
  S observe(S s, X x, Y feedback,
    Map<ModelId, Y> pred);
}
\end{lstlisting}
\vspace{-3em}
\end{figure}


In the current implementation of \system we provide two generic model selection policies based on robust bandit algorithms developed by Auer et al.~\cite{Auer03}.
These algorithms span a trade-off between computation overhead and accuracy.
The single model selection policy (\secref{sec:exp3}) leverages the Exp3 algorithm to optimally \emph{select} the best model based on noisy feedback with minimal computational overhead.
The ensemble model selection policy (\secref{sec:exp4}) is based on the Exp4 algorithm which adaptively \emph{combines} the predictions to improve prediction accuracy and estimate confidence at the expense of increased computational cost from evaluating all models for each query. By implementing model selection policies that provide different cost-accuracy tradeoffs, as well as an API for users to implement their own policies, \system provides a mechanism to easily navigate the tradeoffs between accuracy and computational cost on a per-application basis. Furthermore, users can modify this choice over time as application workloads evolve and resources become more or less constrained.


\subsection{Single Model Selection Policy}
\label{sec:exp3}

We can cast the model-selection process as a multi-armed bandit problem~\cite{Murphy12}.
The multi-armed bandit\footnote{The term bandits refers to pull-lever slot machines found in casinos.}
problem
refers the task of optimally choosing between $k$ possible actions (\eg models) each with a stochastic
reward (\eg feedback).
Because only the reward for the \emph{selected} action can be observed, solutions to the multi-armed bandit problem must address the
trade-off between \emph{exploring} possible actions and \emph{exploiting} the estimated best action.


There are numerous algorithms for the multi-armed bandits problem with a wide range of trade-offs.
In this work we first explore the use of the simple randomized Exp3~\cite{Auer03} algorithm
which makes few assumptions about the problem setting and has strong optimality guarantees.
The Exp3 algorithm associates a weight $s_i = 1$ for each of the $k$ deployed models and then randomly selects model $i$ with probability $p_i = s_i / \sum_{j=1}^k s_j$.
For each prediction $\hat{y}$, \system observes a loss $L(y, \hat{y}) \in [0,1]$ with respect to the true value $y$ (\eg the fraction of words that were transcribed correctly during speech recognition).
The Exp3 algorithm then updates the weight,
$s_i \leftarrow s_i \, \exp\left(-\eta L(y, \hat{y}) / p_i \right)$,
corresponding to the selected model $i$.
The constant $\eta$ determines how quickly \system responds to recent feedback. 


The Exp3 algorithm provides several benefits over manual experimentation and A/B testing, two common ways of performing model-selection in practice.
Exp3 is both simple and robust, scaling well to model selection over a large number of models.
It is a lightweight algorithm that requires only a single model evaluation for each prediction and thus performs well under heavy loads with negligible computational overhead.
And Exp3 has strong theoretical guarantees that ensure it will quickly converge to an optimal solution.

\subsection{Ensemble Model Selection Policies}
\label{sec:exp4}

\begin{table}[t]
\centering
\footnotesize
    \begin{tabular}[b]{ | l | l | c | }
      \hline
      \textbf{Framework} & \textbf{Model} & \textbf{Size (Layers)}  \\
      \hline
      Caffe  & VGG\cite{vgg} &  13 Conv. and 3 FC \\
      Caffe & GoogLeNet\cite{googlenet} &  96 Conv. and 5 FC \\
      Caffe  & ResNet\cite{resnet} &  151 Conv. and 1 FC \\
      Caffe  & CaffeNet\cite{caffenet} & 5 Conv. and 3 FC  \\
      TensorFlow  & Inception\cite{inception} & 6 Conv, 1 FC, \& 3 Incept.\\
      \hline
   \end{tabular}
  \caption{\small \textbf{Deep Learning Models.} The set of deep learning models used to evaluate the ImageNet ensemble selection policy.}
  \label{tab:cnns}
  \vspace{-4mm}
\end{table}

\begin{figure}[t]
\centering \small
\includegraphics[width=\columnwidth]{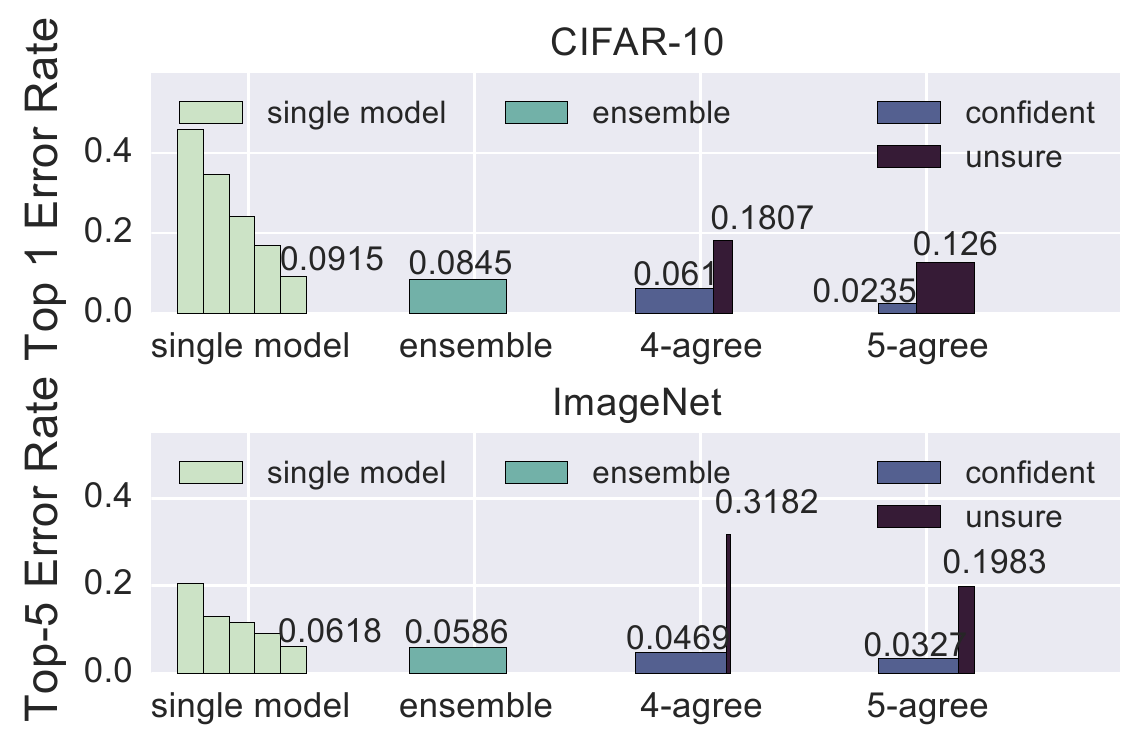}
\caption{\small \textbf{Ensemble Prediction Accuracy.} The linear ensembles are composed of five computer vision models~(\tableref{tab:cnns}) applied to the CIFAR and ImageNet benchmarks.
The 4-agree and 5-agree groups correspond to ensemble predictions in which the queries have been separated
by the ensemble prediction confidence (four or five models agree) and the width of each bar defines the proportion of examples in that category.
}
\label{fig:robust-prediction}
\end{figure}

It is a well-known result in machine learning~\cite{Bishop06,Murphy12,Chen:2016wj,Hinton:2015wy} that prediction accuracy can be improved by combining predictions from multiple models. 
For example, bootstrap aggregation~\cite{Breiman96} (a.k.a., bagging) is used widely to reduce variance and thereby improve generalization performance.  
More recently, ensembles were used to win the Netflix challenge~\cite{Sill:dqm}, and a carefully crafted ensemble of deep neural networks was used to achieve state-of-the-art accuracy on the speech recognition corpus Google uses to power their acoustic models~\cite{Hinton:2015wy}.
The ensemble model selection policies adaptively combine the predictions from \emph{all} available models to improve accuracy, rather than select individual models.


In \system we use linear ensemble methods which compute a weighted average of the base model predictions.
In \figref{fig:robust-prediction}, we show the prediction error rate of linear ensembles on two benchmarks.
In both cases linear ensembles are able to marginally reduce the overall error rate.
In the ImageNet benchmark, the ensemble formulation achieves a 5.2\% relative reduction in the error rate simply by combining off-the-shelf models (\tableref{tab:cnns}). While this may seem small, on the difficult computer vision tasks for which these models are used, a lot of time and energy is spent trying to achieve even small reductions in error, and marginal improvements are considered significant~\cite{imagenet}.

There are many methods for estimating the ensemble weights including linear regression, boosting~\cite{Murphy12}, and bandit formulations.
We adopt the bandits approach and use the Exp4 algorithm~\cite{Auer03} to learn the weights.
Unlike Exp3, Exp4 constructs a weighted \emph{combination} of all base model predictions and updates weights based on the individual model prediction error.
Exp4 confers many of the same theoretical guarantees as Exp3.
But while the accuracy when using Exp3 is bounded by the accuracy of the single best model, Exp4 can further improve prediction accuracy as the number of models increases. The extent to which accuracy increases depends on the relative accuracies of the set of base models, as well as the independence of their predictions.
This increased accuracy comes at the cost of increased computational resources consumed by each prediction in order to evaluate all the base models.



The accuracy of a deployed model can silently degrade over time. \system's online selection policies can automatically detect these failures using feedback and compensate by switching to another model (Exp3) or down-weighting the failing model (Exp4). To evaluate how quickly and effectively the model selection policies react in the presence of changes in model accuracy, we simulated a severe model degradation while receiving real-time feedback.
Using the CIFAR dataset we trained five different Caffe models with varying levels of accuracy to perform object recognition.
During a simulated run of 20K sequential queries with immediate feedback, we degraded the accuracy of the best-performing model after 5K queries and then allowed the model to recover after 10K queries.

\begin{figure}[t]
\centering \small
\includegraphics[width=0.95\linewidth]{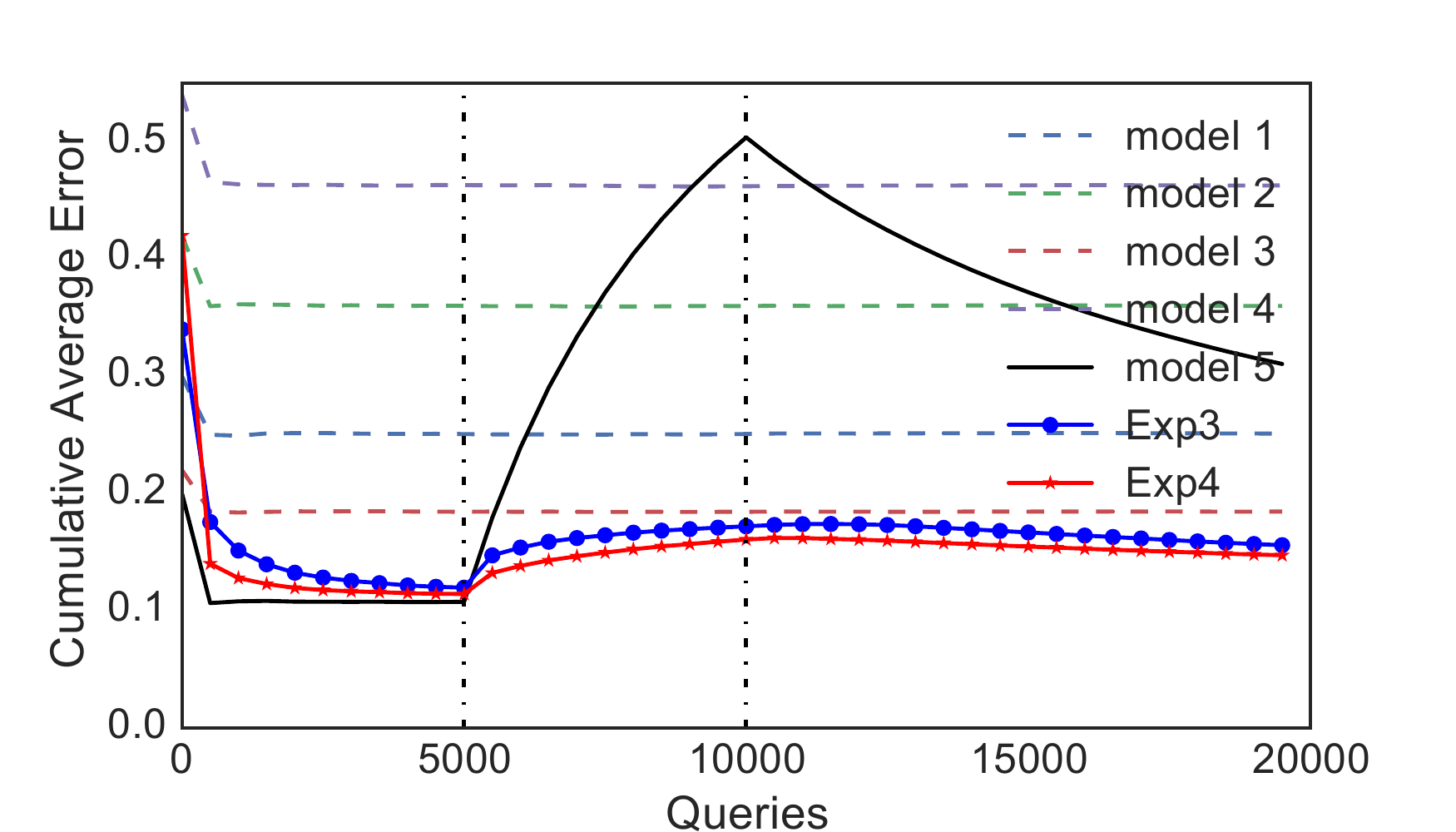}
\vspace{-1em}
\caption{\small \textbf{Behavior of Exp3 and Exp4 Under Model Failure.}
  After 5K queries the performance of the lowest-error model is severely degraded, and after 10k queries performance recovers. Exp3 and Exp4 quickly compensate for the failure and achieve lower error than any static model selection.
}
\label{fig:exp3-eval}
\vspace{-1em}
\end{figure}

In \figref{fig:exp3-eval} we plot the cumulative average error rate for each of the five base models as well as the single (Exp3) and ensemble (Exp4) model selection policies.
In the first 5K queries both model selection policies quickly converge to an error rate near the best performing model (model 5).
When we degrade the predictions from model 5 its cumulative error rate spikes.
The model selection policies are able to quickly mitigate the consequences of the increase in errors by learning to divert queries to the other models.
When model 5 recovers after 10K queries the model selection policies also begin to improve by gradually sending queries back to model 5.

\begin{figure*}[t]
\centering \small
\subfloat[][Latency]{
  \includegraphics[width=0.32\linewidth]{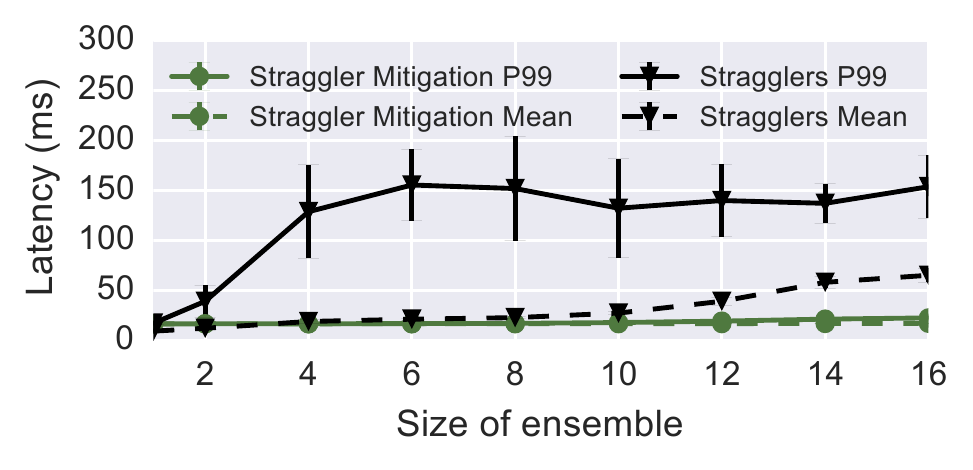}
  \label{fig:stragmit:latency}
}
\subfloat[][Missing Predictions]{
  \includegraphics[width=0.32\linewidth]{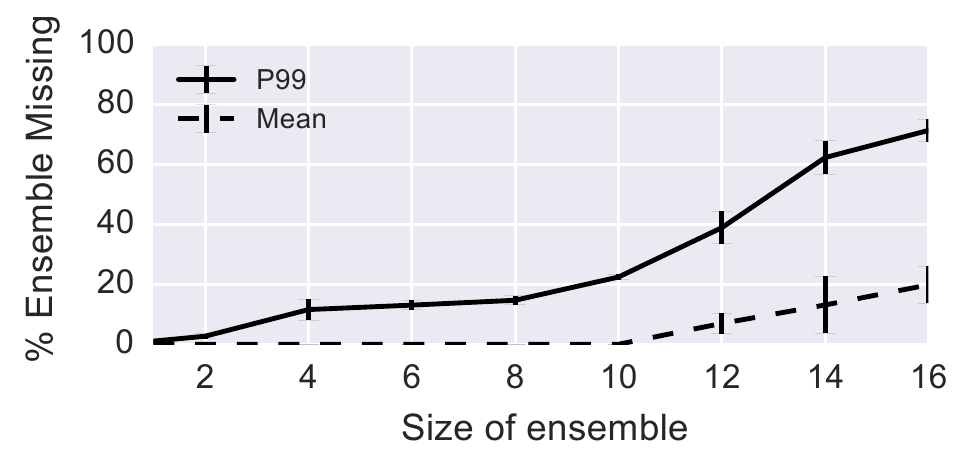}
  \label{fig:stragmit:missing}
}
\subfloat[][Accuracy]{
  \includegraphics[width=0.32\linewidth]{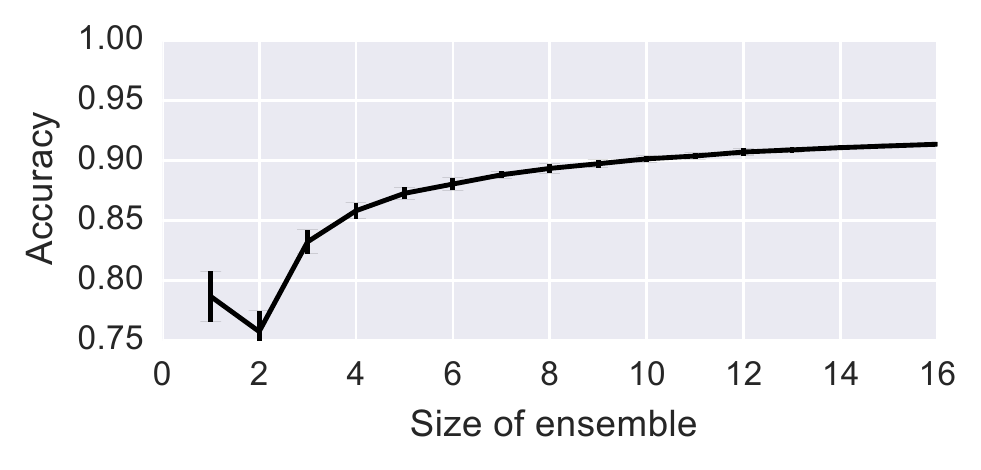}
  \label{fig:stragmit:accuracy}
}
\caption{\small \textbf{Increase in stragglers from bigger ensembles.}
  The \textbf{(a)} latency, \textbf{(b)} percentage of missing predictions, and \textbf{(c)} prediction accuracy when using the ensemble model selection policy on SK-Learn Random Forest models applied to MNIST.
As the size of an ensemble grows, the prediction accuracy increases but the latency cost of blocking until all predictions are available grows substantially. Instead, \system enforces bounded latency predictions and transforms the latency cost of waiting for stragglers into a reduction in accuracy from using a smaller ensemble.}
\label{fig:straggler-mitigation}
\end{figure*}


\subsubsection{Robust Predictions}
\label{subsec:robust-predictions}

The advantages of online model selection go beyond detecting and mitigating model failures to leveraging new opportunities to improve application accuracy and performance. For many real-time decision-making applications, knowing the confidence of the prediction can significantly improve the end-user experience of the application.

For example, in many settings, applications have a sensible default action they can take when a prediction is unavailable. This is critical for building highly available applications that can survive partial system failures or when building applications where a mistake can be costly. Rather than blindly using all predictions regardless of the confidence in the result, applications can choose to only accept predictions above a confidence threshold by using the robust model selection policy.
When the confidence in a prediction for a query falls below the confidence threshold, the application can instead use the sensible default decision for the query and avoid a costly mistake.


By evaluating predictions from multiple competing models concurrently we can obtain an estimator of the confidence in our predictions.
In settings where models have high variance or are trained on random samples from the training data (\eg bagging), agreement in model predictions is an indicator of prediction confidence.
When evaluating the \emph{combine} function in the ensemble selection policy we compute a measure of confidence by calculating the number of models that agree with the final prediction.
End user applications can use this confidence score to decide whether to rely on the prediction.
If we only consider predictions where multiple models agree, we can substantially reduce the error rate (see \figref{fig:robust-prediction}) while declining to predict a small fraction of queries.





\subsubsection{Straggler Mitigation}
\label{subsec:straggler-mitigation}

While the ensemble model selection policy can improve prediction accuracy and help quantify uncertainty, it introduces additional system costs.
As we increase the size of the ensemble the computational cost of rendering a prediction increases.
Fortunately, we can compensate for the increased prediction cost by scaling-out the model abstraction layer.
Unfortunately, as we add model containers we increase the chance of stragglers adversely affecting tail latencies.


To evaluate the cost of stragglers, we deployed ensembles of increasing size and measured the resulting prediction latency (\figref{fig:stragmit:latency}) under moderate query load.
Even with small ensembles we observe the effect of stragglers on the P99 tail latency, which rise sharply to well beyond the 20ms latency objective.
As the size of the ensemble increases and the system becomes more heavily loaded, stragglers begin to affect the mean latency.

To address stragglers, \system introduces a simple best-effort straggler-mitigation strategy motivated by the design choice that rendering a \emph{late} prediction is worse than rendering an \emph{inaccurate} prediction.
For each query the model selection layer maintains a latency deadline determined by the latency SLO.
At the latency deadline the \emph{combine} function of the model selection policy is invoked with the \emph{subset} of the predictions that are available.
The model selection policy must render a final prediction using only the available base model predictions and communicate the potential loss in accuracy in its confidence score.
Currently, we substitute missing predictions with their average value and define the confidence as the fraction of models that agree on the prediction.


The best-effort straggler-mitigation strategy prevents model container tail latencies from propagating to front-end applications by maintaining the latency objective as additional models are deployed.
However, the straggler mitigation strategy reduces the size of the ensemble.
In \figref{fig:stragmit:missing} we plot the reduction in ensemble size and find that
while tail latencies increase significantly with even small ensembles, most of the predictions arrive by the latency deadline.
In \figref{fig:stragmit:accuracy} we plot the effect of ensemble size on accuracy and observe that this ensemble can tolerate the loss of small numbers of component models with only a slight reduction in accuracy.




\subsection{Contextualization}
\label{subsec:contextualization}

\begin{figure}[t]
\centering \small
\includegraphics[width=0.95\linewidth]{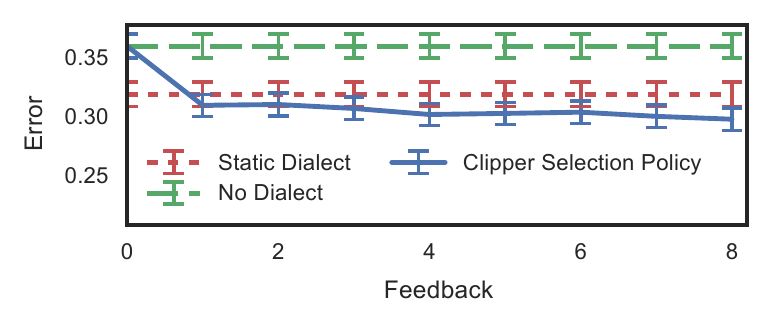}
\vspace{-1.5em}
\caption{\small \textbf{Personalized Model Selection.} Accuracy of the ensemble selection policy on the speech recognition benchmark.
}
\vspace{-1.5em}
\label{fig:cold-start-timit}
\end{figure}

In many prediction tasks the accuracy of a particular model may depend heavily on context.
For example, in speech recognition a model trained for one dialect may perform well for some users and poorly for others.
However, selecting the right model or composition of models can be difficult and is best accomplished online in the model selection layer through feedback.
To support context specific model selection, the model selection layer can be configured to instantiate a unique model selection state for each user, context, or session.
The context specific session state is managed in an external database system.
In our current implementation we use Redis.

To demonstrate the potential gains from personalized model selection we hosted a collection of TIMIT~\cite{timit} voice recognition models each trained for a different dialect.
We then evaluated (\figref{fig:cold-start-timit}) the prediction error rates using a single model trained across all dialects, the users' reported dialect model, and the \system ensemble selection policy.
We first observe that the dialect-specific models out-perform the dialect-oblivious model, demonstrating the value of context to improve prediction accuracy.
We also observe that the ensemble selection policy is able to quickly identify a combination of models that out-performs even the users' designated dialect model by using feedback from the serving workload.

%% file: evaluation.tex

\section{System Comparison}
\label{sec:eval}

In addition to the microbenchmarks presented in \secref{sec:model-abstraction} and \secref{sec:selection-layer}, we compared \system's performance to TensorFlow Serving and evaluate latency and throughput on three object recognition benchmarks.


TensorFlow Serving~\cite{tfserving} is a recently released prediction serving system created by Google to accompany their TensorFlow machine learning training framework.
Similar to \system, TensorFlow Serving is designed for serving machine learning models in production environments and provides a high-performance prediction API to simplify deploying new algorithms and experimenting with new models without modifying frontend applications.
TensorFlow Serving supports general TensorFlow models with GPU acceleration through direct integration with the TensorFlow machine learning framework
and tightly couples the model and serving components in the same process.

TensorFlow Serving also employs batching to accelerate prediction serving.
Batch sizes in TensorFlow Serving are static and rely on a purely timeout based mechanism
to avoid starvation.
TensorFlow Serving does not explicitly incorporate prediction latency objectives which must be achieved by manually tuning the batch size.
Furthermore, TensorFlow Serving was designed to serve one model at a time and therefore does not directly support feedback, dynamic model selection, or composition.



To better understand the performance overheads introduced by \system's layered architecture and decoupled model containers, we compared the serving performance of \system and TensorFlow Serving on three TensorFlow object recognition deep networks of varying computational cost: a 4-layer convolutional neural network trained on the MNIST dataset~\cite{tftutorial}, the 8-layer AlexNet~\cite{alexnet} architecture trained on CIFAR-10~\cite{cifardata}, and Google's 22-layer Inception-v3 network~\cite{inception} trained on ImageNet.
We implemented two \system model containers for each TensorFlow model, one that calls TensorFlow from the more standard and widely used Python API and one that calls TensorFlow from the more efficient C++ API.
All models were run on a GPU using hand-tuned batch sizes (MNIST: 512, CIFAR: 128, ImageNet: 16) to maximize the throughput of TensorFlow Serving.
The serving workload measured the maximum sustained throughput and corresponding prediction latency for each system.

\begin{figure}[t]
\centering
\includegraphics[width=0.95\linewidth]{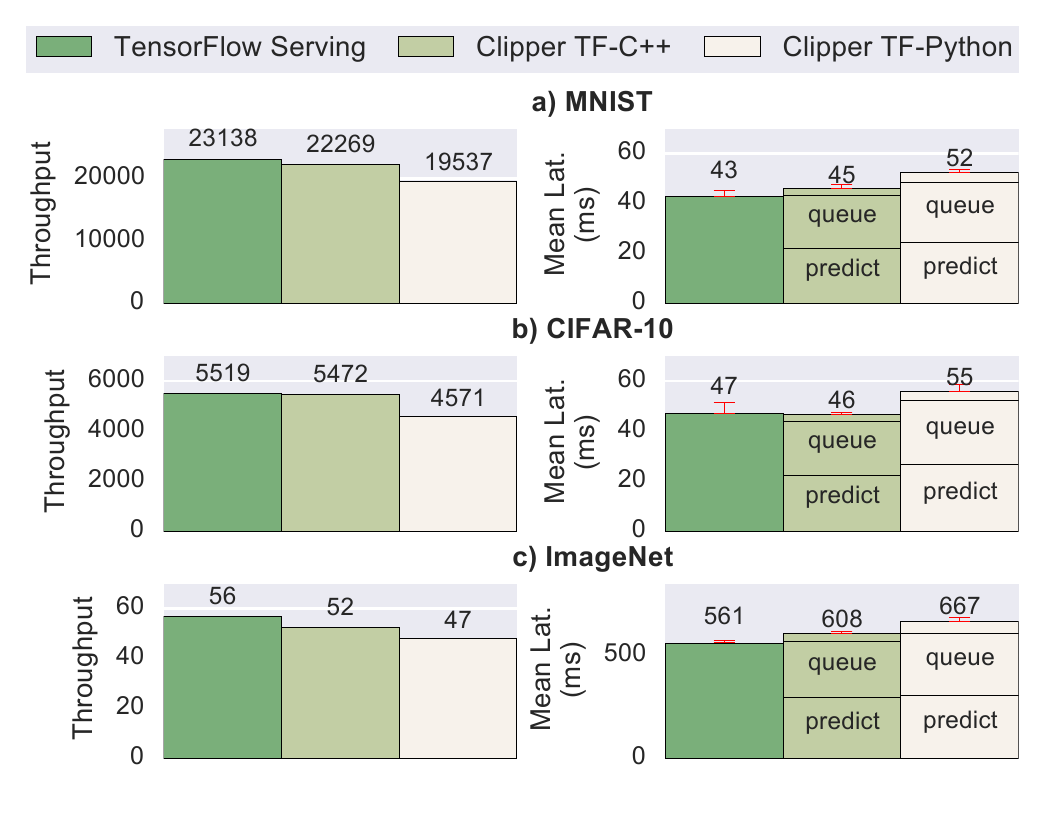}
\vspace{-1em}
\caption{\small \textbf{TensorFlow Serving Comparison.} Comparison of peak throughput and latency (p99 latencies shown in error bars) on three TensorFlow models of varying inference cost. TF-C++ uses TensorFlow's C++ API and TF-Python the Python API.
}
\label{fig:tf_comp}
\vspace{-1em}
\end{figure}

Despite \system's modular design, we are able to achieve comparable throughput to TensorFlow Serving across all three models (\figref{fig:tf_comp}).
The Python model containers suffer a 15-18\% performance hit compared to the throughput of TensorFlow Serving, but the C++ model containers achieve nearly identical performance.
This suggests that the high-level Python API for TensorFlow imposes a significant performance cost in the context of low-latency prediction-serving but that \system does not impose any additional performance degradation.

For these serving workloads, the throughput bottleneck is inference on the GPU. Both systems utilize additional queuing in order to saturate the GPU and therefore maximize throughput. For the \system model containers, we decomposed the prediction latency into component functions to demonstrate the overhead of the modular system design. The \emph{predict} bar is the time spent performing inference within TensorFlow framework code. The \emph{queue} bar is time spent queued within the model container waiting for the GPU to become available. The top bar includes the remaining system overhead, including query serialization and deserialization as well as copying into and out of the network stack. As~\figref{fig:tf_comp} illustrates, the RPC overheads are minimal on these workloads and the next prediction batch is queued as soon as the current batch is dispatched to the GPU for inference. TensorFlow Serving utilizes a similar queueing method to saturate the GPU, but because of the tight integration between TensorFlow Serving and the TensorFlow inference code, they are able to push the queueing into the TensorFlow framework code itself running in the same process.

By achieving comparable performance across this range of models, we have demonstrated that through careful design and implementation of the system, the modular architecture and substantially broader set of features in \system do not come at a cost of reduced performance on core prediction-serving tasks.

\section{Limitations}

While \system attempts to address many challenges in the context of prediction serving there are a few key limitations when compared to other designs like TensorFlow Serving.
Most of these limitations follow directly from the design of the \system architecture which assumes models are below \system in the software stack, and thus are treated as black-box components.

\system does not optimize the execution of the models within their respective machine learning frameworks.
Slow models will remain slow when served from \system.
In contrast, TensorFlow Serving is tightly integrated with model evaluation, and hence is able to leverage GPU acceleration and compilation techniques to speedup inference on models created with TensorFlow.

Similarly, \system does not manage the training or re-training of the base models within their respective frameworks.
As a consequence, if all models are out-of-date or inaccurate \system will be unable to improve accuracy beyond what can be accomplished through ensembles.



%% file: related-work.tex

\section{Related Work}
\label{sec:related-work}

The closest projects to \system are LASER~\cite{Agarwal14}, Velox~\cite{Crankshaw15}, and TensorFlow Serving~\cite{tfserving}.
The LASER system was developed at LinkedIn to support linear models for ad-targeting applications.
Velox is a UC Berkeley research project to study personalized prediction serving with Apache Spark.
TensorFlow Serving is the open-source prediction serving system developed by Google for TensorFlow models.
In our experiments we only compare against TensorFlow Serving, because LASER is not publicly available, and the current prototype of Velox has very limited functionality.

All three systems propose mechanisms to address latency and throughput.
Both LASER and Velox utilize caching at various levels in their systems.
In addition, LASER also uses a straggler mitigation strategy to address slow feature evaluation.
Neither LASER or Velox
discuss batching.
Conversely, TensorFlow Serving does not employ caching and instead leverages batching and hardware acceleration to improve throughput.

LASER and Velox both exploit a form of model decomposition to incorporate feedback and context
similar to the linear ensembles in \system.
However, LASER does not incorporate feedback in real-time, Velox does not support bandits and neither system supports cross framework learning.
Moreover, the techniques used for online learning and contextualization in both of these systems are captured in the more general \system selection policy.
In contrast, TensorFlow Serving has no mechanism to achieve personalization or  adapt to real-time feedback.

Finally, LASER, Velox, and TensorFlow Serving are all vertically integrated; they focused on serving predictions from a single model or framework.
In contrast, \system supports a wide range of machine learning models and frameworks and simultaneously addresses latency, throughput, and accuracy in a single serving system.


\textbf{Application Specific Prediction Serving:}
There has been considerable prior work in application and model specific prediction-serving.
Much of this work has focused on content recommendation, including video-recommendation~\cite{Davidson:2010hg},
ad-targeting~\cite{McMahan:2013cq,Graepel10}, and product-recommendations~\cite{Lerallut:2015gg}.
Outside of content recommendation, there has been recent success in speech recognition~\cite{Lei:2013wy,siri} and internet-scale resource allocation~\cite{Ganjam:2015tq}.
While many of these applications require real-time predictions, the solutions described are highly application-specific and tightly coupled to the model and workload characteristics.
As a consequence, much of this work solves the same systems challenges in different application areas.
In contrast, \system is a general-purpose system capable of serving many of these applications.

\textbf{Parameter Server:}
There has been considerable work in the learning systems community on parameter-servers~\cite{Dean12paramserver,MuLi,Ahmed12,Xing15}.
While parameter-servers do focus on reduced latency and caching, they do so in the context of \emph{model training}.
In particular they are a specialized type of key-value store used to coordinate updates to model parameters in a distributed training system.
They are not typically used to serve predictions.




\textbf{General Serving Systems:} The high-performance serving architecture of \system draws from prior work on highly-concurrent serving systems\cite{Pai:1999ws,Welsh:2001kv,Schmidt:1995uq,nginx}. The division of functionality into vertical stages introduced by~\cite{Welsh:2001kv} is similar to the division of \system's architecture into independent layers.
Notably, while the dominant cost in data-serving systems tends to be IO, in prediction serving it is computation. This changes both physical resource allocation and batching and latency-hiding strategies.

%% file: conclusion.tex
\section{Conclusion}


In this work we identified three key challenges of prediction serving: latency, throughput, and accuracy, and proposed a new layered architecture that addresses these challenges by interposing between end-user applications and existing machine learning frameworks.

As an instantiation of this architecture, we introduced the \system prediction serving system.
\system isolates end-user applications from the variability and diversity in machine learning frameworks by providing a common prediction interface.
As a consequence, new machine learning frameworks and  models can be  introduced without modifying end-user applications.

We addressed the challenges of prediction serving latency and throughput within the \system Model Abstraction layer.
The model abstraction layer lifts caching and adaptive batching strategies above the machine learning frameworks to achieve up to a 26x improvement in throughput while maintaining strict bounds on tail latency and providing mechanisms to scale serving across a cluster.
We addressed the challenges of accuracy in the \system Model Selection Layer.
The model selection layer enables many models to be deployed concurrently and then dynamically selects and combines predictions from each model to render more robust, accurate, and contextualized predictions while mitigating the cost of stragglers.



We evaluated \system using four standard machine-learning benchmark datasets
spanning computer vision and speech recognition applications.
We demonstrated \system's capacity to bound latency, scale heavy workloads across nodes, and provide accurate, robust, and contextual predictions.
We compared \system to Google's TensorFlow Serving system and achieved parity on throughput and latency performance, demonstrating that the modular container-based architecture and substantial additional functionality in \system can be achieved with minimal performance penalty.